\documentclass[sigconf]{acmart}
\renewcommand\footnotetextcopyrightpermission[1]{} 
\setcopyright{none}
\usepackage{booktabs} 
\usepackage{hyperref}
\usepackage[ruled]{algorithm2e} 

\setcopyright{none}

\begin{document}
\title[On the Feasibility of Real-Time 3D Hand Tracking using Edge GPGPU Acceleration]{On the Feasibility of Real-Time 3D Hand Tracking\\ using Edge GPGPU Acceleration}
 
\author{Ammar Qammaz}
\affiliation{%
  \institution{FORTH and University of Crete}
  \city{Heraklion}
  \state{Crete}
  \country{Greece}}
\email{ammarkov@ics.forth.gr} 

\author{Sokol Kosta}
\affiliation{%
  \institution{Aalborg University Copenhagen}
  \city{Copenhagen}
  \state{Denmark}
}
\email{sok@es.aau.dk}

\author{Nikolaos Kyriazis}
\affiliation{%
  \institution{Insitute of Computer Science, FORTH}
  \streetaddress{N. Plastira 100, Vassilika Vouton,}
  \city{Heraklion}
  \state{Crete}
  \postcode{GR70013}
  \country{Greece}}
\email{nkyriazis@gmail.com}

\author{Antonis Argyros}
\affiliation{%
  \institution{FORTH and University of Crete}
  \city{Heraklion}
  \state{Crete}
  \country{Greece}}
\email{argyros@ics.forth.gr}

\renewcommand\shortauthors{A. Qammaz et al.}

\begin{abstract}
This paper presents the case study of a non-intrusive porting of a monolithic C++ library for real-time 3D hand tracking, to the domain of edge-based computation. Towards a proof of concept, the case study considers a pair of workstations, a computationally powerful and a computationally weak one. By wrapping the C++ library in Java container and by capitalizing on a Java-based offloading infrastructure that supports both CPU and GPGPU computations, we are able to establish automatically the required server-client workflow that best addresses the resource allocation problem in the effort to execute from the weak workstation. As a result, the weak workstation can perform well at the task, despite lacking the sufficient hardware to do the required computations locally. This is achieved by offloading computations which rely on GPGPU, to the powerful workstation, across the network that connects them. We show the edge-based computation challenges associated with the information flow of the ported algorithm, demonstrate how we cope with them, and identify what needs to be improved for achieving even better performance.
\end{abstract}

%
%

%
%


\maketitle

\section{Introduction}
   Mobile devices have become ubiquitous in western societies, networking technologies are constantly improving, and cloud computing services offer an abundance of resources for applications that support them.
   On the other hand, factors such as cost of purchase, power consumption, battery life, as well as the large number of legacy mobile devices still being used by consumers, make a large portion of the mobile computer market underpowered relative to the top-tier devices commercially available. 
    Of course, high-end devices such as gaming laptops do exist and offer amazing performance, but their cost is prohibitive for the average consumer and they mainly cater to a smaller gamer niche. This leaves most devices equipped with cheap hardware alternatives such as on-board graphics and slower and more power conserving CPUs. This, in turn, translates to less computing resources for applications albeit in more affordable computers.
   Another trend, which is one of the motivations behind this work, is that computer users increasingly view the world from their devices with a surprising number of new HCI devices recently entering the market, mostly in the form of virtual assistants such as Amazon Echo, Google Home, Siri, and Kortana, that currently lack vision and rely on voice commands and only provide audio output using a network back-end. 
\begin{figure}
  \includegraphics[height=90pt]{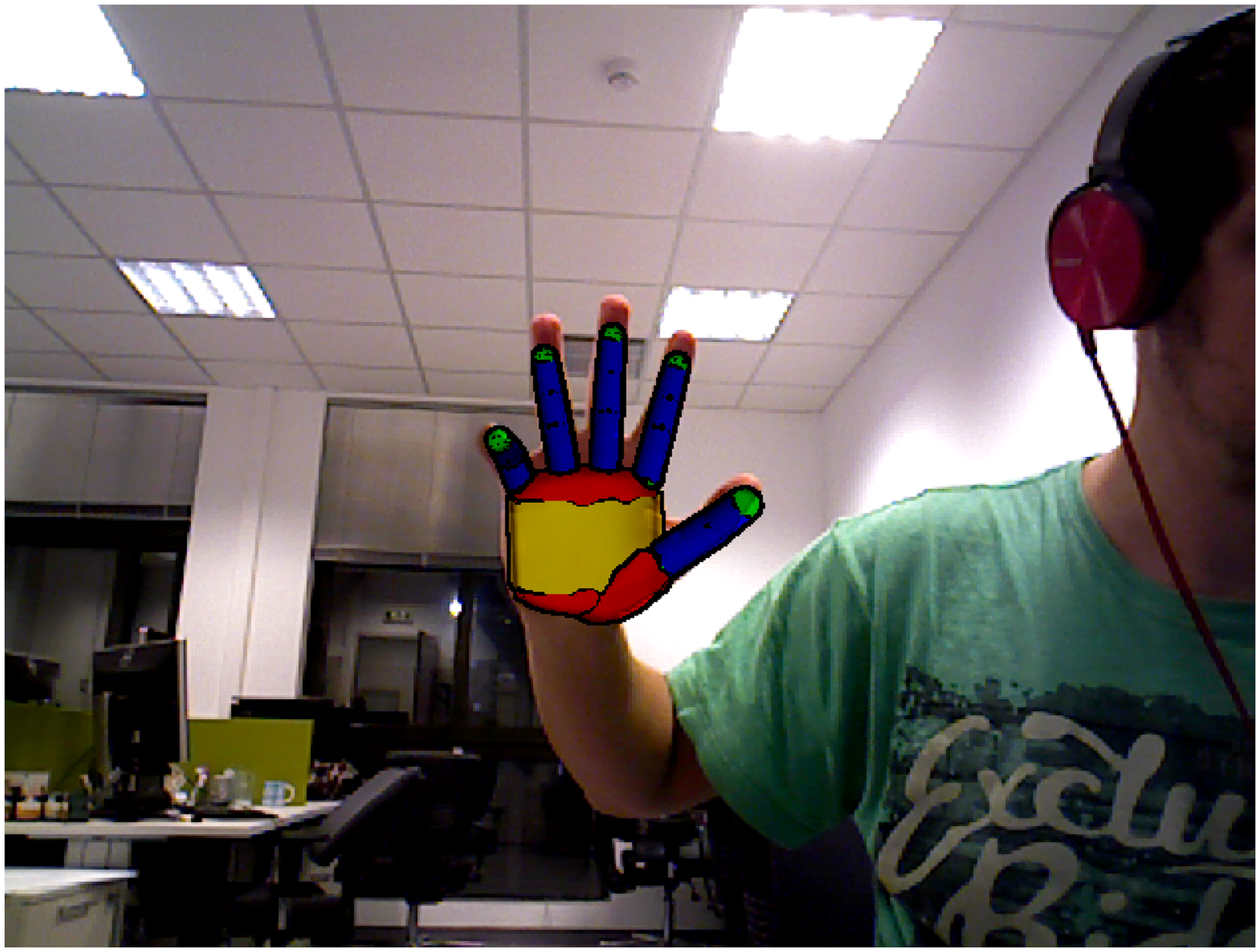}  \includegraphics[height=90pt]{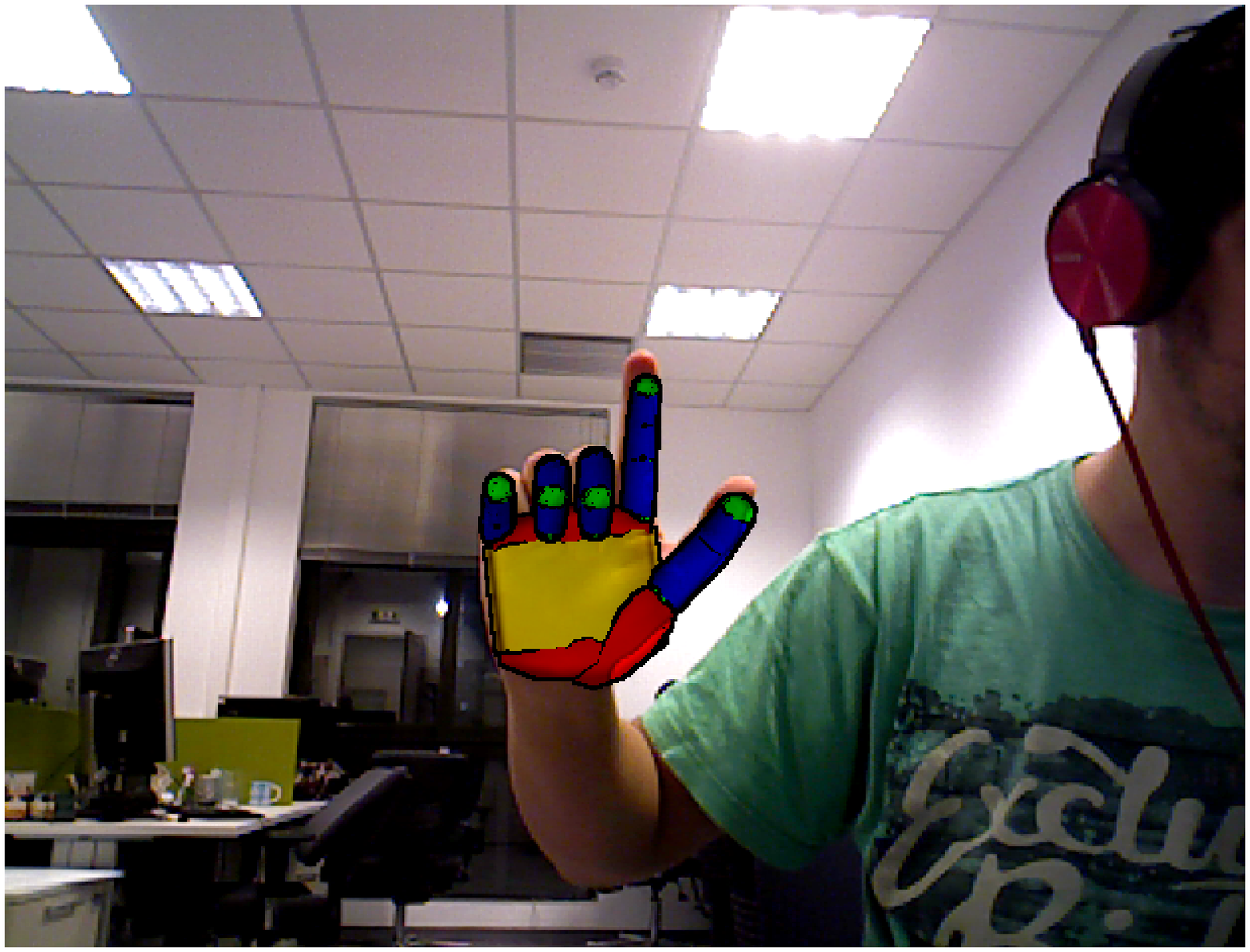}
  \caption{Indicative hand tracker output overlayed over the RGB feed of an RGBD camera source. 
  }
  \label{fig:visualization}
\end{figure}
In a world that is becoming increasingly connected, we believe that humans will rely more and more on computer vision techniques to facilitate human-computer interaction.
These trends hint that this technology brings great future potential for applications that can cater to the needs of users by leveraging the tools currently available.
As an example, a robust, vision-based hand tracker module enables a computer system to recover the 3D position, orientation and full articulation of a human hand unobtrusively, from visual input. This is a stepping stone for building additional components such as gesture recognition. If this functionality becomes accessible to low-end devices, a variety of tasks (as simple as controlling a media player or as complex as household robotics and managing smart-homes by using hand signs), can be performed via simple, intuitive gestures.

In this work, we report on the experience and challenges of extending a powerful generative real-time 3D Hand Tracker to make it exploit Edge Offloading for achieving satisfactory results even when running on low-power devices.

  The rest of the paper is organized as follows: 
  Section~\ref{sec:related} presents an overview of the most relevant works in the field;
  Section~\ref{sec:system} describes the details of the GPU-based implementation of the hand tracking system and its adaption for low-power devices thanks to Edge Offloading;
  Section~\ref{sec:experiments} presents the results of the experiments performed in different configuration setups; 
  and finally, Section~\ref{sec:conclusions} concludes the paper with final remarks on the lessons learned by running a real-time application on the edge via computation offloading.

\section{Related work}\label{sec:related}

We review the research works that are most relevant to real-time hand tracking and network acceleration/computation offloading.

\subsection{Hand Tracking}
 Real-Time 3D Hand Tracking is an open research problem in Computer Vision. Several approaches have been proposed towards achieving an effective and efficient solution. A very recent overview of such research efforts and state of the art hand tracking solutions is presented in~\cite{Yuan2018}. Methods can be roughly categorized based on the camera sensors they use and on the adopted hand pose estimation approach. These approaches are broadly classified into discriminative, generative, and hybrid. 
Recent discriminative works use Convolutional Neural Networks (CNNs) with off-the-shelf RGB cameras but as seen in~\cite{Yuan2018}, 3D volumetric representations outperform 2D CNNs and 2D CNNs are not directly suitable for 3D hand pose estimation~\cite{ge20173d}. 
We thus focus on real-time methods that use RGBD/depth sensors. Discriminative methods such as \cite{wan2017crossing} use statistical modeling, other works~\cite {ge20173d,oberweger2017deeppriorplus} use 3D CNNs, while other solutions use cascades~\cite{sun2015cascaded} or forest based classifiers~\cite{tang2014latent}. 
Generative approaches to the problem involve the use of particle filters~\cite{MakrisKyriazisArgyros2015a} or genetic algorithms~\cite{OikonomidisKyriazisArgyros2011b} to perform regression and solve the problem.
Hybrid solutions with both discriminative and generative components have also been proposed~\cite{krejov2017guided,sharp2015accurate}.
It is important to note that almost all state of the art methods utilize GPGPU acceleration in order to achieve high framerates.

\subsection{Automatic Network Acceleration}
Computation offloading is a simple but powerful concept, which has been used successfully to help thin client devices execute resource and energy-hungry applications~\cite{Xu2018}.
The first proposed solutions, such as Cloudlets~\cite{Satyanarayanan2009}, MAUI~\cite{Cuervo2010}, CloneCloud~\cite{Chun2011} and ThinkAir~\cite{Kosta2012}, among others, have shown that it is beneficial for low-powered devices to offload the heavy tasks to remote powerful entities for remote execution, under the right circumstances. Indeed, these works show that the computation offloading works well when at least one or more of the following conditions are satisfied: 
\textit{i)} the task to be offloaded is computationally heavy, \textit{ii)} the amount of data to be transferred is limited, and \textit{iii)} the network between the thin client and the powerful remote entity is good.
For this reason, the applications considered by these works can be classified as CPU-hungry, delay-tolerant (up to a certain level, meaning they are not real-time), and not data-intensive.

With the recent progress in the distribution of networking, storage, and computing resources, i.e. increasing number of data centers around the world, and with the advent of the 5G network, more advanced offloading frameworks have been proposed, which deal with applications that present also challenging features, such as real-time or data-intensive. Flores et al.~\cite{Flores2017} present a framework for computation offloading from Internet of Things (IoT) devices, with a solution for load-balancing on the cloud side when handling multiple requests for task offloading.
Chatzopoulos et al.~\cite{chatzopoulos2016openrp} propose OpenRP, a platform for avoiding selfish behavior in the context of Device-to-Device offloading, where mobile devices can help each-other execute tasks in a distributed manner.

Montella et al.~\cite{Montella:2014} are the pioneers of open source solutions for GPU code offloading in low power devices.
RAPID\footnote{\url{http://www.rapid-project.eu/}} is a European project that provides a framework for CPU and GPU computation offloading from low-powered devices to the edge of the network or to the cloud, depending on the context~\cite{Montella2016,Montella2017}. In the current work, we make use of the RAPID tools to implement and test the GPU-based real-time hand tracking on thin clients.


\section{3D Hand Tracking on the Edge}\label{sec:system}

\subsection{Generative 3D Hand Tracking}
\begin{figure}[t]
  \includegraphics[width=\columnwidth]{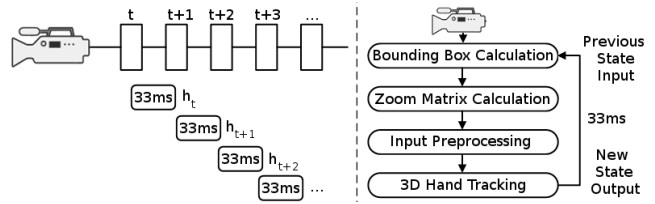}
  \caption{Left: A real-time framerate of 30fps means that every frame acquired by a camera has to be consumed/processed in less than 33 milliseconds. Right: 3D hand tracking consists of four discrete optimization steps which need to be executed in the 33ms time window. We use a JNI container that exposes these steps to the offloading framework and can choose to offload each of them separately (Multiple-Step experiments, Figure~\ref{fig:performance}) or all of them in a single step (Single-Step experiments, Figure~\ref{fig:performance}) in order to reduce intermediate data transfers and network latency.   }
  \label{fig:archIdeal}
\end{figure}

The hand tracker application is a processor of frames generated by a camera at a framerate of 30 frames per second (fps). It can be considered as a black box that receives a hand configuration $h_t$ at time $t$ (hand 3D position, orientation, articulation), along with a frame pair of RGB $c^o$ and depth $d^o$ images at time $t+1$ and computes an estimation $h_{t+1}$ of the hand configuration for the input frame. Repeating the procedure for every acquired RGBD frame results in the tracking of the observed hand. 

\vspace*{0.15cm}
{\bf \noindent Hand model:} 
Each hand is encoded as a vector $h$ of $27$ kinematic parameters. $3$ parameters represent its 3D location, and $4$ more parameters represent its 3D orientation using a right-hand Cartesian coordinate system $(X, Y, Z)$ and using a quaternion representation to avoid gimbal locks. The rest $20$ parameters represent bone angles that encode finger articulation. 
 
\vspace*{0.15cm}
{\bf \noindent Objective  function:}
An RGBD frame is denoted as $o=(c^o, d^o)$, where $c^o$ and $d^o$ stand for the camera RGB and depth frames, respectively.
Moreover, given an instance $h$ of the hand model and the camera calibration parameters, we can render the hand model to the camera viewport, obtaining color and depth maps $r=(c^h, d^h)$ that are directly comparable to the observations.
An objective function $E_D(h,o)$ can then be defined that quantifies the discrepancy between a hand configuration / model hypothesis $h$ and the actual observations $o$. Estimating the hand pose at a certain frame amounts to finding the model parameters $h^*$ that minimize $E_D(h,o)$:
\begin{equation}
h^* \buildrel \Delta \over = \mathop {\arg \min }\limits_h E_D\left( {h,o} \right).
\label{eq:of}
\end{equation} 
More specifically, the error term that quantifies the discrepancy between a hand configuration $h$ and the available observations, $o$ is defined as:
\begin{equation}
E_D\left(h, d^o\right) = \frac{1}{N_P} {\sum_{p \in B}  C \left( |d^h_p - d^o_p|, T \right)}.
\label{eq:depthterm}
\end{equation}
In Eq.(\ref{eq:depthterm}), $E_D$  sums the absolute depth differences $|d^h_p - d^o_p|$ for all points $p$ that belong to a bounding box $B$ containing the hand. The clamping function $C(x,T)$ returns $x$ if $x \leq T$ and $T$ otherwise. This is used to make the error term robust and prevent outliers from affecting it too much. In our implementation we set $T=30$cm. 

Hand pose estimation in a certain frame can thus be thought of as an optimization/minimization problem that seeks for the best set of $27$ parameter values that result in a hand pose hypothesis that best matches the available observations. 

\vspace*{0.15cm}
{\bf \noindent Particle Swarm Optimization (PSO):}
\label{pso} 
Particle Swarm Optimization (PSO)~\cite{clerc2002particle} is a stochastic method that performs optimization by iteratively improving a candidate solution with respect to an error term characterizing its quality (objective function). 
PSO maintains a population of candidate solutions, called particles, that have a position and a velocity in the search space. The movement of each particle is influenced by the best position this particle has ever visited up to the current iteration/generation, and simultaneously guided towards the globally best known position in the search space (i.e., the best of the best for all particles). This strategy moves the swarm toward the best solutions.

PSO does not require training and does not need to compute the gradient of the optimized objective function which, thus, does not need to be differentiable. PSO is ideal for parallel implementation on modern GPU architectures, permitting interactive framerates and a GPGPU implementation provides $100\times$ speedup compared to a serial implementation.

PSO has been applied successfully to a number of vision problems such as object detection~\cite{StefanouArgyros2012a}, head pose estimation~\cite{PadelerisZabulisArgyros2012a}, 3D hand tracking~\cite{OikonomidisKyriazisArgyros2011b,OikonomidisKyriazisArgyros2012a}, 3D tracking of hands in interaction with objects~\cite{KyriazisArgyros2014a,PantelerisKyriazisArgyros2015a,QammazKyriazisArgyros2015a}, estimating contact forces~\cite{PhamKheddarQammazEtAl2015}, as well as 3D human pose tracking~\cite{vijay2010,MichelPanagiotakisArgyros2015a,ammar2018}.
Tracking more complex object configurations than a single hand increases the search space of PSO exponentially. As an example, tracking two hands results in a $54$-D search space, and each subsequent rigid object adds $7$ more dimensions (its position plus its orientation)~\cite{KyriazisArgyros2014a,QammazKyriazisArgyros2015a}. In turn, the increase of the dimensionality of the problem creates the need for more computational resources and a larger optimization budget (i.e., number of evaluations of candidate solutions). Offloading to powerful machines, with more computational resources than a thin client, is expected to help towards an effective solution to this problem.
In this work, we examine the basic $27$ degree of freedom (dof) single hand-tracking problem~\cite{OikonomidisKyriazisArgyros2011b}, since it represents the first step towards addressing the challenges of real-time computation offloading in the edge for facilitating Human-Computer Interaction (HCI) and other real-life scenarios.

\vspace*{0.15cm}
{\bf \noindent PSO for 3D hand tracking:} For our 3D hand tracking problem, PSO minimizes the objective function of Eq.(\ref{eq:of}) over candidate solutions $h$. For each incoming frame, particles are initialized around the solution of the previous frame. The space around that solution is made large enough to include the current frame estimation.  

\begin{figure}[tb!]
  \includegraphics[width=\columnwidth]{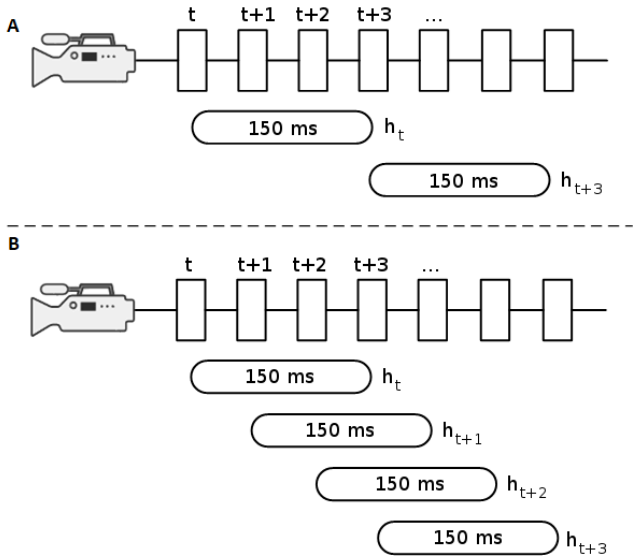}
  \caption{A: Algorithms with serial frame dependencies have to wait until a frame is processed before consuming the next. B: An offloading architecture can accommodate much better rates when not having to wait for a request to finish. Generative 3D Hand Tracking falls in category A.}
  \label{fig:archSerial}
\end{figure}

The optimization internally happens in 4 consecutive steps, as depicted in Figure~\ref{fig:archIdeal}.
In order to achieve a 30 fps rate, and process every frame in real-time, the system should perform every processing step in less than 33 ms. Otherwise, if the system takes more than 33 ms per step, the whole pipeline gets slowed down, because the hand configuration $h_{t+1}$ cannot be estimated without considering the estimation $h_t$ of the hand in the previous time step.
This is depicted in segment A of Figure~\ref{fig:archSerial}, where we can observe that for a hypothetical slower 150 ms processing loop time, the system must skip processing two consecutive frames for each received frame, since it does not have enough processing time for them. 
Dropping frames is bad not only for the user experience (as there is observable delay), but also for the quality of the tracking. 
During the time unaccounted for, the hand moves further away from the last tracked/estimated position. This means that the hand tracker has to compensate by considering hand configurations in a much wider area in the space of hand poses, which makes the problem much more difficult. Additionally, errors tend to accumulate and can quickly lead to track loss in case of fast motion and low frame rates.

This also means that we cannot simply transmit a video feed over the network and just collect tracking results at the rate of acquisition. Instead, our client needs to wait for a frame to be processed and then return back with the best explanation $h$ in order to submit the next one and start a new search. This makes it a particularly hard case since we are aiming at real-time performance but any latencies experienced through the network medium are aggregated to the overall loop time of the client.

\subsection{3D Hand Tracking on Thin Clients} 

We use the RAPID platform as the offloading mechanism for enabling real-time hand tracking on thin clients. RAPID is a framework for Java and Android low-power devices, which can be used for both CPU and GPGPU code offloading.
As such, given that the Hand Tracker was originally written as a C++ library, we created a Java Native Interface (JNI) container for the library, as well as a Java front-end that allowed us to achieve automatic offloading using RAPID.
RAPID's use of the maven build system allow effortless packaging of all the required runtime-libs of the application across machines, and the Java front-end makes it very easy to control and script it on a high-level without exposing the developer to the complexity of tens of thousands lines of code underneath.

When the RAPID-enabled application starts for the first time, a registration process is triggered, which involves sending the bytecode of the application to the remote side of the RAPID system.
Being a method-level offloading framework, whenever an offloadable Java method is called, RAPID decides at runtime if the method should be executed locally on the device or remotely on the powerful machine. If the offloaded Java method uses \textit{native} C/C++ functions, which is the case of our application, the RAPID remote counterpart will dynamically load the shared libraries (\emph{.so} files) embedded in the application, which were transmitted during the registration phase, and will try to resolve the native function. Thanks to this feature, we were able to easily to utilize the offloading framework straightforwardly.
Moreover, if a native function contains also CUDA calls, which again is the case of our application, they are automatically executed on the GPU as if they were running normally without any offloading involved.

Thanks to offloading, we are not constrained by the resources of a single machine, which could potentially allow us to scale up as seen in Figure~\ref{fig:archSerial} on part B of the diagram.
Unfortunately, as described in the previous sections, the nature of the optimization framework used for the hand tracker is not suited for this kind of parallel processing, given that each frame must first be processed and produce the output before the next one can be processed. Thus, we lose the benefits of concurrency and any potential benefit gained by assigning each incoming frame to a separate computing resource that can accommodate it, given that the system needs to wait for each step to be completed before handling the next step.

The source-code of the Java/JNI wrapper implementation is distributed as open source and can be found in GitHub~\cite{github}. The C++ Hand Tracker native library is also provided in binary form in the repository, allowing for the reproduction of the following experiments.

\section{Experimental Evaluation}\label{sec:experiments}
\subsection{Experiment Setup}

In order to study the behavior of our RAPID-enabled hand tracker application, we use a testing environment with two different tiers of devices: a high-end server and a low-end laptop. 
The high-end server features a GeForce GTX 1080M GPU and an Intel Core i7 processor, while the laptop has an outdated GeForce 670M GPU and an early generation Intel Core i5.

Moreover, we connect the devices using two networks: a fast Gigabit Ethernet connection and a slower 802.11 Wi-Fi connection.

As mentioned above, to achieve a 30 fps tracking loop rate, which allows to process every frame received from the RGBD device, all the processing needs to be executed within 33 ms.
Unfortunately, Wi-Fi connections are very prone to radio interference and typically introduce latency ranging from 10--60 ms, depending on the number of connected clients and network saturation. Moreover, the available bandwidth of a Wi-Fi connection is substantially lower than the one provided by a Gigabit Ethernet connection.

In order to have comparable results with the different setups, we pre-recorded a video depicting various challenging hand movements.

Having the same input stream to evaluate across all runs, we aim to identify the overhead introduced by the network connections, by the RAPID offloading framework, and by the Java's JNI wrapper and the Java Virtual Machine (JVM).

\subsection{Quantitative Results}
To help with reading the results presented in this section, Table~\ref{tab:definitions} provides a brief summary of the main terminology used in the description of the experiments.

\begin{figure}[!tb]
  \includegraphics[height=180pt]{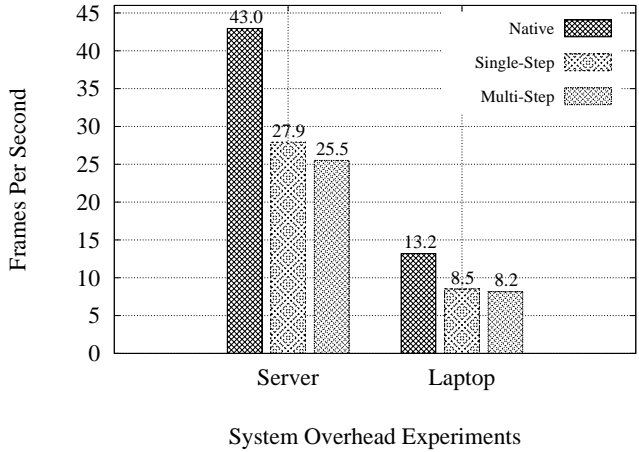}
  \caption{Performance graph of the native C++ implementation of the hand tracker in the two devices \textit{vs.} the RAPID-enabled Java version without offloading.
  The server presents better results than the laptop, due to the outdated hardware of the laptop.
  The overhead added by the offloading framework is less pronounced in the laptop, due to the overall slower framerate achieved on the older hardware.}

  \label{fig:performanceSys}
\end{figure}

We begin our evaluation by measuring the sustainable performance of the Hand Tracker in its vanilla non-Java, non-RAPID implementation, when executed in the high-end server and the low-end laptop respectively. The results of this analysis are the baseline of our evaluation and are displayed in Figure~\ref{fig:performanceSys}. The high-end hardware available in the server computer allows the C++ native application to achieve real-time processing at a rate bigger than 40 frames per second.
The 30 fps limit is important since it matches the rate at which the RGBD camera acquires new Depth and RGB frames.
On the other hand, the slower laptop can only achieve an average rate of 13 fps for the same C++ implementation, which is much smaller.

Then, we proceed by measuring the performance of the RAPID-enabled implementation of the hand tracker, when executed on the server and laptop host respectively, without utilizing code offloading. The results of these experiments are again portrayed in Figure~\ref{fig:performanceSys}. This analysis enables us to identify the overhead introduced by wrapping the native code of the Hand Tracker inside a Java container using JNI. The results obtained at this step reveal the impact of data serialization, synchronization, and JVM overheads. We observe that in this configuration, the application's performance is reduced when executed on the high-end server host. When executing on the low-end laptop host, where the potential GPU speedup is lower and the overall execution speed slower, the overhead introduced by the Java Wrapper is much less evident.
The results also show that when wrapping each tracking step individually in Java methods (the \textit{Multi-Step} case), the overhead is more visible compared to having all the steps in a single Java method (the \textit{Single-Step} case).

\begin{figure}[!tb]
 \begin{center}
  \includegraphics[height=180pt]{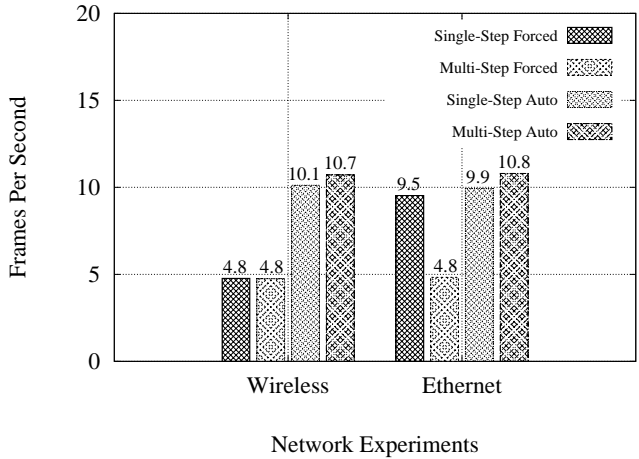}
  \caption{Performance graph of the offloaded hand tracking from the laptop to the server with Wi-Fi and Ethernet under different configurations.
  The \textit{Forced} configuration presents better results when the hand tracking steps are implemented in a single Java method (Single-Step) and when Ethernet is used. The \textit{Auto} configuration presents same results in all situations.
  }

  \label{fig:performance}
  \end{center}
\end{figure}

\begin{table}[!tb]
\centering
\caption{Terminology used in the experimental setup and related results.}
\label{tab:definitions}
\begin{tabular}{rp{6.3cm}}
\toprule                                                                               \textbf{Server} & High-end machine with a GeForce GTX 1080M GPU and an Intel Core i7 processor. \\
\textbf{Laptop} & Thin-client laptop with an outdated GeForce 670M GPU and an early generation Intel Core i5. \\
\textbf{Native} & The C++ hand tracking implementation. \\
\textbf{Single-Step} & The hand tracking steps presented in Figure~\ref{fig:archIdeal} are called inside a single Java method. \\
\textbf{Multi-Step} & The hand tracking steps presented in Figure~\ref{fig:archIdeal} are called via separate Java methods. \\
\textbf{Forced} & RAPID is instructed to always offload an offloadable method. This is used to measure the performance in the case of a thin-client without GPU, which needs to always offload. \\
\textbf{Auto} & RAPID decides automatically whether to offload or not an offloadable method. This is used when the thin-client can run the method also locally, because it has a GPU. \\
\bottomrule
\end{tabular}
\end{table}

Finally, we measure the performance characteristics of the distributed hand tracker, utilizing code offloading via the Gigabit Ethernet and the Wi-Fi connection. The purpose of this study is to evaluate the performance gain obtained by offloading the heavy parts of the application logic from the low-end laptop to the high-end server host.

When RAPID is instructed to always offload the offloadable methods (the \textit{Forced} case), which would happen in the case of a device without a GPU, the \textit{Single-Step} implementation under the Ethernet connection yields the best performance, with a framerate around 10 fps.
This means that thanks to offloading, a machine without a GPU is possible to run the real-time 3D hand tracking with $1/3$ of the desired framerate. Even though this might not be acceptable for a commercial use, it is still a good start towards building better systems that can improve these results.

If the thin-client is able to run the tasks also locally, such as in our case that the laptop has a GPU, RAPID can dynamically decide for each task to run it either locally or remotely. These results, which are noted as \textit{Auto} in the graphs of Figure~\ref{fig:performance}, show that RAPID is able to adapt in all situations and yield the best possible performance even if the connection is Wi-Fi rather than Ethernet. Also in this case, the achieved framerate is around $10-11$ fps.

\section{Conclusions and Future Work}\label{sec:conclusions}

In this work, we presented the first results of the successful adaptation of a GPU-based real-time 3D hand tracker for executing it in thin-clients thanks to the computation offloading paradigm.
We showed that, thanks to the maturity of the existing offloading frameworks, the effort of transforming the C++ native application to a distributed version was minimal.

Our experiments showed that choosing a Java-based offloading framework might not be ideal when it comes to real-time or near real-time applications. Indeed, our evaluations confirmed that the overhead of the Java layer is not negligible, and it considerably reduced the performance of the application in terms of framerate.
However, our experiments proved that Edge Computation offloading can help limited low-power devices execute applications that cannot be run otherwise because of lack of resources, such as the GPU in our hand tracking case.
 
As future work, we intend to work on two axes to improve the results presented here: \textit{i)} investigate on the possibility of using or developing a real-time oriented offloading framework, so that applications will not be penalized by its overhead, and \textit{ii)} investigate single-frame 3D hand pose estimators such as the one described in~\cite{Panteleris2018}. This could allow for a higher level of parallelization, since there would be no inter-frame dependencies. In such case, all newly acquired frames could be submitted directly to the computing resources without any stall, the network delay would not be accumulated, and offloading could provide a substantial improvement of the resulting computational performance.

\begin{acks}
The work was partially supported by the European Commission Horizon 2020 programs RAPID  (H2020-ICT-644312) and Co4Robots (H2020-ICT-2016-1-73186). 
\end{acks}

\bibliographystyle{IEEEtran}
\bibliography{main}

\end{document}